\documentstyle[12pt]{article}

\newlength{\dinwidth}
\newlength{\dinmargin}
\newfont\goth{eufm10 scaled 1200}
\newfont\bbl{msbm10 scaled 1095}
\newfont\bbs{msbm10 scaled 1000}
\newfont\bbss{msbm9 scaled 1000}
\newfont\bbsss{msbm7 scaled 1000}

\def\Z{\mbox{\bbl Z}}

\def\1{{\bf 1}}

\def\ga{\mbox{$\g(A)$}}

\newtheorem{satz}{Proposition}

\def\beq{\begin{equation}}      \def\eeq{\end{equation}}
\def\baro{\begin{eqnarray*}}    \def\barr{\begin{eqnarray}}
\def\earo{\end{eqnarray*}}      \def\earr{\end{eqnarray}}
\newcommand{\bensatz}{\begin{satz} {\bf:} \begin{enumerate}}
\newcommand{\bsatz}{\begin{satz} {\bf:}\\ }
\newcommand{\esatz}{\end{satz}}
\newcommand{\beenden}{\end{enumerate}}
\newcommand{\beginnen}{\begin{enumerate}}

\newcommand{\al}{\alpha}
\newcommand{\sbf}[4]{\!{\textstyle \left[ {{#1}\atop {#2}}{{#3}\atop
    {#4}} \right] }}
\newcommand{\be}{\begin{equation}}
\newcommand{\eq}{\end{equation}}
\newcommand{\bz}{\bar{z}}
\newcommand{\ra}{\rightarrow}
\newcommand{\CF}{{\cal F}}
\newcommand{\CG}{{\cal G}}
\newcommand{\D}{\Delta}
\newcommand{\BB}{\bar{B}}
\newcommand{\Ga}{\Gamma}
\renewcommand{\ga}{\gamma}
\newcommand{\Up}{\Upsilon}
\newcommand{\ket}[1]{| #1 \rangle}
\newcommand{\bra}[1]{\langle #1 |}
\begin{document}
\thispagestyle{empty}
\renewcommand{\thefootnote}{\fnsymbol{footnote}}
\vspace*{2cm}
\begin{center}
{\LARGE \sc On the Liouville three-point function \\[3.5mm]
}
\vspace*{1cm}
            {\sl J\"org Teschner\footnote[3]{Supportet by Deutsche
        Forschungsgemeinschaft}}\\
 \vspace*{6mm}
     2. Institut f\"ur theoretische Physik, Universit\"at Hamburg\\
     Luruper Chaussee 149, D-22761 Hamburg, Germany\\
 \vspace*{6mm}
     July 20th, 1995\\
\vspace*{1cm}
\begin{minipage}{11cm}\footnotesize
The recently proposed expression for the general
three point function of exponential fields in quantum Liouville theory
on the sphere is considered.
By exploiting locality or crossing symmetry in the case
of those four-point functions, which may be expressed in terms of
hypergeometric functions, a set of functional equations is found for the
general three point function. It is shown that the expression proposed
by the Zamolodchikovs solves these functional equations and that
under certain assumptions the solution is unique.
\end{minipage}
\end{center}
\renewcommand{\thefootnote}{\arabic{footnote}}
\setcounter{footnote}{0}
\newpage
Up to now exact information on exponential Liouville operators and their
correlation functions was available only for a rather small subclass
of these operators. On a heuristic level these results were based on
the observation that the Liouville path integral with exponential operators
inserted in some cases reduces to a path integral in a free field
theory \cite{GL}. These rather wild manipulations can be rigorously
justified within the context of the exact operator quantization as
developed by Gervais and collaborators \cite{GN1}\cite{GN2}\cite{GS}.
However, these methods only work in the case that the number of
screening charges in a free field representation of operators and
correlation functions is a positive integer. There have been attempts
to extend these results to the case of a negative integer number of
screening charges, based either on continuation prescriptions
\cite{GL}\cite{Do}\cite{Ki}\cite{FK} or on conjectures about the chiral
algebra of the corresponding chiral vertex operators \cite{Ge2}, but
rigorous justification for these proposals is still missing.\par
Recently a proposal was made for the
exact general three point function of exponential Liouville field
operators (\cite{DO1}\cite{DO2} and independently in \cite{ZZ}).
If properly established, such a result could be an important progress
beyond the known cases. It provides the information needed for the
decomposition of arbitrary correlation functions into conformal blocks
(for the general formalism see \cite{BPZ}, a proposal in the context
of Liouville theory is eqn. (2.23) of \cite{ZZ}), and should therefore
be an important ingredient to study the factorization properties of
Liouville theory. Factorization of Liouville correlation functions is
a subtle issue \cite{Sei}\cite{Pol}, and one may expect important
qualitative insights on the nature of Liouville theory and 2d gravity
from such investigations.\par
The aim of the present note is to show that under certain additional
assumptions the requirement of crossing symmetry \cite{BPZ} may be
used to give a proof for the expression proposed in \cite{ZZ}. \par
Let $V_{\al}$ denote the Liouville field
operator corresponding to the classical field $e^{2\al\phi}$.
Application of these operators on the $sl_2$-invariant
state\footnote{Existence of such a state has been
  questioned in the context of the Gervais-Neveu approach.
  However, in \cite{T} it has been shown that most achievments of
  this approach are indeed compatible with the existence of an
  $sl_2$-invariant state.}
$\ket{0}_L$ creates Virasoro highest weight states $\ket{\al}_L$ with weight
$h_{\al}=\al(Q-\al)$:
\be \lim_{z,\bz\ra\infty}V_{\al}(z,\bz)\ket{0}_L=\ket{\al}_L. \eq
To label Liouville states $\ket{\al}$ by "momenta" $\al$ is justified by
the correspondence to free field states:
The asymptotic behaviour of the wave-functional
$\Psi_{\ket{\al}}(\phi(z,\bz))$ in the asymptotic limit when the
zero-mode $\phi_0$ goes to $-\infty$ is given by the free field wave
functions of highest weight states $\ket{\al}_F$ with zero-mode momentum
$\al$, see i.e. \cite{Pol2} or \cite{ZZ}. One gets the correspondence
\be \ket{\al}_L \quad\sim\quad \ket{\al}_F+R(\al)\ket{Q-\al}_F.
\label{corr}\eq
For imaginary values of $\al$, $\al=iP+Q/2$, one may interprete
$R(\al)$ as the amplitude for reflection on the Liouville wall.\par
The Liouville space of
states is covered once by restricting $\al-Q/2\leq 0$. However, it is
convenient to also introduce states with $\al-Q/2\geq 0$, defined by
\be \ket{Q-\al}_L=R^{-1}(\al)\ket{\al}_L\quad\sim\quad
\ket{Q-\al}_F+R^{-1}(\al)\ket{\al}_F.\eq
Obviously, it is useful to define $R(Q-\al)=R^{-1}(\al)$ in order to
have the correspondence to free field states in the form of (\ref{corr}).
\par
{}From now on I will only consider Liouville states and drop the
subscript $L$.
The states dual to $\ket{\al}$ will be denoted $\bra{\al}$. The
out-vacuum will be defined to be the $sl_2$-invariant state $\bra{Q}$.
I will assume that
\be
\lim_{z,\bz\ra\infty}\bra{Q}V_{\al}(z,\bz)|z|^{4\D_{\al}}=\bra{Q-\al}
\label{outass}\eq
This assumption identifies the reflection amplitude $R(\al)$ with the
two-point function $\langle V_{\al}V_{\al}\rangle$.\par
The three point function is defined as
\be
\CG_{\al_3\al_2\al_1}(x_3,x_2,x_1)=
\bra{Q}V_{\al_3}(z_3,\bz_3)V_{\al_2}(z_2,\bz_2)
V_{\al_1}(z_1,\bz_1)\ket{0}, \eq
where $x_i=(z_i,\bz_i)$.
$V_{\al}$ is supposed to be a primary conformal field with dimension
$\D_{\al}=\al(Q-\al)$. The coordinate dependence is therefore
determined to be \cite{BPZ}
\be \CG_{\al_3\al_2\al_1}(x_3,x_2,x_1)=
|z_{12}|^{\D_{12}}|z_{13}|^{\D_{13}}|z_{23}|^{\D_{23}}
C(\al_3,\al_2,\al_1), \eq
where $z_{ij}=z_i-z_j$, $\D_{ij}=\D_k-\D_i-\D_j$ for $k\neq i,j$ and
$\D_i\equiv\D_{\al_i}$.
The structure constants $C(\al_3,\al_2,\al_1)$ are the main object of
interest in the present note. As a consequence of locality, $C$ has to
be symmetric in its arguments.\par
A consequence of assumption (\ref{outass}) that will be important
below is
\be \bra{\al_3}V_{\al_2}(z,\bz)\ket{\al_1}\propto
C(Q-\al_3,\al_2,\al_1). \label{opecoeff} \eq
Now consider the four point function of exponential Liouville operators,
which will be written as
\be
\CG_{\al_4\al_3\al_2\al_1}(x_4,x_3,x_2,x_1)=
\bra{Q}V_{\al_4}(z_4,\bz_4)V_{\al_3}(z_3,\bz_3)V_{\al_2}(z_2,\bz_2)
V_{\al_1}(z_1,\bz_1)\ket{0},\label{corrdef}
\eq
Projective invariance allows to reduce $\CG$ to a function of the
cross-ratio
\be z=\frac{z_{21}z_{43}}{z_{31}z_{42}} \eq
and its complex conjugate:
\begin{eqnarray}\lefteqn{ \CG_{\al_4\al_3\al_2\al_1}(x_4,x_3,x_2,x_1)=
}\\
& = & |z_{42}|^{-4\D_2}
|z_{41}|^{2(\D_3+\D_2-\D_1-\D_4)}|z_{43}|^{2(\D_1+\D_2-\D_3-\D_4)}
|z_{31}|^{2(\D_4-\D_1-\D_2-\D_3)}
G_{\al_4\al_3\al_2\al_1}(z,\bz).
\nonumber\end{eqnarray}
As a consequence of (\ref{opecoeff}) the decomposition of
$G_{\al_4\al_3\al_2\al_1}(z,\bz)$ into conformal blocks is of the
general form
\be  G_{\al_4\al_3\al_2\al_1}(z,\bz)=
\sum_{\al}C(\al_4,\al_3,\al)C(Q-\al,\al_2,\al_1)
\left|\CF_s\sbf{\al_3}{\al_4}{\al_2}{\al_1}(z)\right|^2.
\label{CBdeco} \eq
An important property of $\CG_{\al_4\al_3\al_2\al_1}(x_4,x_3,x_2,x_1)$
that follows from the requirements of locality and associativity of
the operator products is the so-called crossing-invariance \cite{BPZ},
which in terms of $G_{\al_4\al_3\al_2\al_1}(z,\bz)$ is expressed as
\be
G_{\al_4\al_3\al_2\al_1}(z,\bz)=G_{\al_4\al_1\al_2\al_3}(1-z,1-\bz)
=|z|^{-4\D_2}G_{\al_1\al_3\al_2\al_4}(1/z,1/\bz).
\label{cross} \eq
The strategy to get information on the coupling constants will be to
exploit (\ref{cross}) in a special case where the conformal blocks in
(\ref{CBdeco}) can be determined explicitely:\par
For the special case that $\al_2=-b/2$ (resp. $\al_2=-1/2b$) it is
well known \cite{BPZ}\cite{GN2}\footnote{The relation between these
  rather different approaches is clarified in \cite{T}}
that the operator $V_{\al_2}$
satisfies operator differential equations:
\be \begin{array}{c@{\qquad}c}
\multicolumn{2}{c}{\displaystyle
\frac{\partial^2}{\partial z^2}V_{\al_2}(z,\bz)=
b^2(T_<(z)V_{\al_2}(z,\bz)+V_{\al_2}(z,\bz)T_>(z)),
}\\[.3cm]\displaystyle
T_<(z)=\sum_{n=-\infty}^{-2}z^{-n-2}L_n & \displaystyle
T_>(z)=\sum_{n=-1}^{\infty}z^{-n-2}L_n
\end{array} \eq
and a corresponding equation for the antiholomorphic dependence.
As a consequence, $G$ satisfies the ordinary
differential equations
\be
\left(-\frac{1}{b^2}\frac{d^2}{dz^2}+
  \left(\frac{1}{z-1}+\frac{1}{z}\right)
    \frac{d}{dz}-\frac{\D_3}{(z-1)^2}-\frac{\D_1}{z^2}+
    \frac{\D_3+\D_2+\D_1-\D_4}{z(z-1)} \right)G(z,\bz)=0.
\label{de}\eq
It follows from (\ref{de}) that the only values of $\al$ that appear
in (\ref{CBdeco}) are $\al_1+sb/2$, $s=\pm 1$ and that the conformal blocks
$\CF_s\equiv \CF_{\al_1+sb/2}$ may be expressed in terms of the
hypergeometric function
$F(A,B;C;z)$ as
\be \CF_s(z)=z^{a_s}(1-z)^bF(A_s,B_s;C_s;z)\eq
where $a_s=\D_{\al_1+sb/2}-\D_2-\D_1$, $b=\D_{\al_3-b/2}-\D_3-\D_2$ and
\begin{eqnarray}
A_s &=& -sb(\al_1-Q/2)+b(\al_3+\al_4-b)-1/2 \\
B_s &=& -sb(\al_1-Q/2)+b(\al_3-\al_4)+1/2 \\
C_s &=& 1-sb(2\al_1-Q).
\end{eqnarray}
The basic point here is that the identity
\begin{eqnarray}
F(A,B;C;z) &=&
\frac{\Ga(C)\Ga(B-A)}{\Ga(B)\Ga(C-A)}(-z)^{-A}F(A,1-C+A;1-B+A,1/z)
\nonumber \\
 & &
 \frac{\Ga(C)\Ga(A-B)}{\Ga(A)\Ga(C-B)}(-z)^{-B}F(B,1-C+B;1-A+B,1/z).
\label{hypid} \end{eqnarray}
yields a relation of the form
\be
\CF_s\sbf{\al_3}{\al_4}{\al_2}{\al_1}(z)=z^{-2\D_2}\sum_{t=+,-}\;B_{st}\;
\CF_s\sbf{\al_3}{\al_1}{\al_2}{\al_4}(1/z), \eq
which may be used to exploit the crossing symmetry relations
(\ref{cross}). One finds
\be \frac{C(\al_4,\al_3,\al_1+b/2)}{C(\al_4,\al_3,\al_1-b/2)}
=-\frac{C_-(\al_1)}{C_{+}(\al_1)}\frac{B_{-+}\BB_{--}}{B_{++}\BB_{+-}} ,
\label{funeq}\eq
where the notation $C_s(\al)=C(\al,-b/2,Q-(\al+sb/2))$ has been used.
This is the sought-for functional equation for
$C(\al_4,\al_3,\al_1)$, since the right hand side may be determined
explicitely: \\
First, the explicit form of the $B_{st}$ is found from
(\ref{hypid}). This yields
\begin{eqnarray} \lefteqn{
\frac{B_{-+}\BB_{--}}{B_{++}\BB_{+-}}=
-\frac{\ga(b(2\al_1-b))}{\ga(2-b(2\al_1-b))}\times} \\
& & \times
\frac{\ga(b(-\al_1+\al_3+\al_4-b/2))}{\ga(b(\al_1-b/2+\al_3+\al_4-Q))
\ga(b(\al_1-\al_3+\al_4-b/2))\ga(b(\al_1+\al_3-\al_4-b/2))},\nonumber
\label{bfrac}\end{eqnarray}
where $\ga(z)=\Ga(z)/\Ga(1-z)$ and the identity
\be \frac{\Ga^2(2-z)}{\Ga^2(z)}=-\frac{\ga(2-z)}{\ga(z)} \eq
have been used.\\
Second, the $C_s(\al)$ are among the coupling constants for which
integral representations exist and have been computed in \cite{DF},
cf. also eqn. (3.7) of \cite{ZZ}.
\be \frac{C_-(\al_1)}{C_+(\al_1)}=-\frac{\ga(-b^2)}{\pi\mu}
\ga(2\al_1b)\ga(2-b(2\al_1-b)).
\label{cfrac}\eq
Having found an functional equation for $C(\al_4,\al_3,\al_1)$ one
should compare it with the corresponding equation satisfied by the
expression proposed in \cite{ZZ}, eqn. (3.14). Using the functional
equation for the Upsilon-function,
\be \Up(x+b)=\ga(bx)b^{1-2bx}\Up(x), \eq
one finds
\be \frac{C(\al_3,\al_2,\al_1+b)}{C(\al_3,\al_2,\al_1)}
=-\frac{\ga(-b^2)}{\pi\mu}
\frac{\ga(b(2\al_1+b))\ga(2b\al_1)\ga(b(\al_2+\al_3-\al_1-b))}
{\ga(b(\al_1+\al_2+\al_3-Q))\ga(b(\al_1+\al_2-\al_3))\ga(b(\al_1+\al_3-\al_2))}
,\eq
which is equivalent to (\ref{funeq}) with (\ref{bfrac}),(\ref{cfrac}).\\
One gets a second functional equation by replacing $b\ra b^{-1}$.
The solution to these functional equations is unique up to a constant
factor if one assumes that $b$ is irrational and that the dependence
of $C$ on its last argument is continuous:
If there was a second solution
$D(\al_3,\al_2,\al_1)$ of the same functional equations, then $R=D/C$
would have to be periodic with two incommensurable periods. It is
amusing to note that it was J. Liouville, who first proved that such a
function can only be constant, see \cite{JL}, chap. XIII.
If one expands R in a Fourier-series over trigonometric functions
with period $b$, then every term in the sum has to be
$b^{-1}$-periodic seperately, which is possible only if $R=const.$.
Note that one needs only square-integrability of $R$ in ${[}0,b{]}$ to
reach this conclusion. \par
A number of points deserve further discussion:
\begin{enumerate}
\item The condition of irrationality of $b$ does not hold in the case
  of the minimal models and the corresponding $c\geq 25$ Liouville
  theories needed for their coupling to gravity. The corresponding
  arbitrariness in the solution for $C$ may be fixed by demanding
  something like continuous dependence on the parameter
  $b$. Furthermore, if $1\leq c\leq 25$ then the argument for
  uniqueness breaks down completely since $b$ becomes complex. It is
  intriguing to see how the so-called $(c=1)$-barrier appears in this
  context.
\item A closely related reasoning may be used in an operator approach
  such as that of Gervais and collaborators: First, the notations are
  related by
\newcommand{\bg}{\sqrt{\frac{h}{\pi}}}
\newcommand{\vp}{\varpi}
\be b\equiv \bg\qquad\qquad \al-Q/2\equiv -\frac{1}{2}\bg\vp. \eq
Second, the four point functions to be considered read
\be
\bra{\vp_4}e^{-jb\phi}(1,1)e^{-\frac{1}{2}b\phi}(z,\bz)\ket{\vp_1},
\eq
where
\be e^{-\frac{1}{2}b\phi}(z,\bz)=\sum_{i=1,2}C_{i}(\vp)V_i(z)\bar{V}_i(\bz),
\eq
the chiral components $V_i(z)$ being characterized by the operator
differential equation
\be
\frac{d^2}{dz^2}V_{i}(z)=
\frac{h}{\pi}(T_<(z)V_{i}(z)+V_{i}(z)T_>(z)).
\eq
Now the argument is somewhat dependent on the form that one assumes
for the operator $e^{-jb\phi}(1,1)$: Consider the general ansatz
\be
e^{-jb\phi}(z,\bz)=\int\!\! dr\; C_r^j(\vp)V_r^j(z)\bar{V}_r^j(\bz),
\label{expans}\eq
where $V_r^j(z)$ shifts the momentum by $\vp\ra\vp+2j-2r$, such that
$r$ may formally be identified with the number of
screening charges. The structure constants $C_r^j(\vp)$ correspond to
$C(Q-(\al-jb+rb),-jb,\al)$ in my previous notation.
The requirement of mutual locality of
$e^{-jb\phi}(1,1)$ and $e^{-\frac{1}{2}b\phi}(z,\bz)$ leads to the
same functional equation as above, which now is of the form
\be \frac{C^j_r(\vp_1-1)}{C^j_{r+1}(\vp_1+1)}
=-\frac{C_-(\vp_1)}{C_{+}(\vp_1)}\frac{B_{-+}\BB_{--}}{B_{++}\BB_{+-}},
\label{funeq2}\eq
If one assumes
$C_r^j(\vp)$ to be a square-integrable function of $r,j,\vp$ then the
previous argument for the uniqueness of the solution of the functional
equation goes through. However, if the sum in the ansatz
(\ref{expans}) for $e^{-jb\phi}(z,\bz)$ includes only a discrete
number of values of $r$ then there is no reason to assume that
$C_r^j(\vp)$ depends square-integrably on $r$. Nevertheless, if a value
$r_0$ appears in the sum, then the sum will also have to include
a discrete series $\{r_0+n\}\subset r_0+\Z$, which is restricted only
by possible zeroes or poles of the rhs of (\ref{funeq2}).
Equation (\ref{funeq2}) may then be used for a recursive determination
of $C_{r_0+n}^j(\vp)$ in terms of $C_{r_0}^j(\vp)$. For $r_0=0$ this
will reproduce the results of \cite{DF} on the structure constants as
well as the Goulian-Li continuation prescription.
\item A puzzling issue is the following: In
  \cite{Sei}\cite{Pol}\cite{ZZ} it is proposed that the sum over
  intermediate states mainly includes macroscopic states (hyperbolic
  sector), perhaps with a discrete sum of contributions of microscopic
  states (elliptic sector). However, in the present case $\al_2=-b/2$
  one has only
  microscopic intermediate states. This is probably related to a
  point noted in \cite{ZZ}: If poles cross the contour of integration
  in eqn. (2.23) of \cite{ZZ} then one will get additional discrete
  contributions. Maybe in some cases the integration contour can be
  closed to yield only a finite sum of residues, which can be
  identified with contributions of microscopic states.\\
If the proposal
of \cite{Sei}\cite{Pol}\cite{ZZ} is right, then the ansatz for a
general exponential Liouville operator would have to read
\be  e^{-jb\phi}(z,\bz)=\sum_{r} C_r^j(\vp)V_r^j(z)\bar{V}_r^j(\bz)
+\int\! dP \; C^j(P,\vp)V^j(P,z)\bar{V}^j(P,\bz),
\eq
where $V^j(P,z)$ maps a state from the elliptic sector into the
hyperbolic sector. An
understanding of these issues will be a major advance in Liouville
theory.
\end{enumerate}
\newcommand{\CMP}[3]{{\it Comm. Math. Phys. }{\bf #1} (#2) #3}
\newcommand{\NP}[3]{{\it Nucl. Phys. }{\bf B#1} (#2) #3}
\newcommand{\PL}[3]{{\it Phys. Lett. }{\bf B#1} (#2) #3}
\newcommand{\MPL}[3]{{\it Mod. Phys. Lett. }{\bf #1} (#2) #3}
\newcommand{\PRL}[3]{{\it Phys. Rev. Lett. }{\bf #1} (#2) #3}
\newcommand{\PTPS}[3]{{\it Progr. Theor. Phys. Suppl. }{\bf #1} (#2) #3}

\end{document}